\documentclass[preprint,aps]{revtex4}
\usepackage{amsmath}

\begin{document}

\title{Radiation Reaction: General approach and applications, especially to
electrodynamics}

\author{R. F. O'Connell}
\affiliation{Department of Physics and Astronomy, Louisiana State University, Baton
Rouge, LA 70803-4001 USA}
\date{\today }

\begin{abstract}
Radiation reaction (but, more generally, fluctuations and dissipation)
occurs when a system interacts with a heat bath, a particular case being the
interaction of an electron with the radiation field. We have developed a
general theory for the case of a quantum particle in a general potential
(but, in more detail, an oscillator potential) coupled to an arbitrary heat
bath at arbitrary temperature, and in an external time-dependent $c$-number field. The results may be applied to a large variety
of problems in physics but we concentrate by showing in  detail the application to the blackbody radiation
heat bath, giving an exact result for radiation reaction problem which has no unsatisfactory features such as the runaway solutions associated with the Abraham-Lorentz theory.  In addition, we show how atomic energy and free energy shifts due to temperature may be calculated.  Finally, we give a brief review of applications to Josephson junctions, quantum statistical mechanics, mesoscopic physics, quantum information, noise in gravitational wave detectors, Unruh radiation and the violation of the quantum regression theorem.
\end{abstract}

\maketitle

\section{Introduction}

Radiation reaction is familiar to most readers through the Abraham-Lorentz
equation for a radiating electron \cite{abraham02,lorentz16,jackson98}. It
arises from the fact that electric and magnetic fields emitted by an
accelerating electron act back on the electron, resulting in a retarding
force. However, the equation obtained by Abraham \cite{abraham02} and Lorentz
\cite{lorentz16} displays unphysical runaway solutions. Through the years,
there have been many attempts to obtain
a more satisfactory result but none were without difficulties of one kind or
another until in 1991 we proposed a physically consistent solution in the
form of a quantum Langevin equation \cite{ford91}. We discuss that result
below in Sec. III, but first we consider more general properties of
dissipation and fluctuations, since they appear in many areas of physics.

The equation of motion of a quantum particle in a heat bath (also referred
to as a reservoir or as the environment) was pioneered in 1965 by Ford, Kac
and Mazur \cite{ford65}. These authors used a quantum model of coupled
oscillators to obtain a quantum Langevin equation that showed explicitly the
role of fluctuation and radiation reaction (dissipation) effects. By 1983,
there was much general interest in mesoscopic systems and Caldeira and
Leggett \cite{caldeira83} showed that quantum heat baths could play an
important role in the analysis of such systems by analyzing in detail the
effect of dissipation on quantum tunneling in such systems as a single
Josephson junction or a SQUID. These authors, again considering an Ohmic
heat bath, used the path integral influence functional approach of Feynman
and Vernon \cite{feynman63}, a technique which was then followed by many
investigators, particularly these working on coherence and entanglement
problems. By contrast, the present authors, in collaboration with J. T.
Lewis, developed a general quantum Langevin equation approach to these
problems \cite{ford85,ford88} that, in our opinion, is simpler and more physically
transparent. In particular, we went beyond the Ohmic model to consider an
arbitrary frequency-dependent heat bath (in particular a blackbody radiation
field heat bath).

In general, coupling a system to a heat bath produces two related effects:
dissipation and fluctuations. The study of these phenomena is loosely
referred to as the "Brownian motion,"
problem, going back to the work of Robert Brown, a Scottish botanist who in
1828 and 1829 published his work on the random motion of pollen grains
immersed in a fluid \cite{brown1829}. The explanation of Brown's phenomenon
was given by Einstein and Smoluchowski in the first decade of the twentieth
century \cite{einstein56,smoluchowski}. This work helped to definitively
establish the atomic theory of matter since the irregular motion was clearly
identified as being due to collisions with the molecules in the liquid. The
term "Brownian motion," is now used in a
generic sense to denote random motion and it covers a wide spectrum of
phenomena from the motion of very fine particles suspended in a gas to the
motion of electrons in a blackbody radiation heat bath.

Einstein's explanation of Brownian motion used a discrete time approach. In
particular, his results included an explanation of the fact that increasing
temperatures lead to more agitated Brownian motion, in a relation in which
the diffusion coefficient is shown to be equal to the product of $kT$ and
the mobility, perhaps the first example of what is now called the
fluctuation-dissipation theorem.

Shortly after the work of Einstein and Smoluchowski, Langevin \cite
{langevin08} presented a different approach to the subject which, in the
words of Chandrasekhar \cite{chandra43}, constitutes the "
modern," approach to this and other such problems. The
essence of \emph{Langevin's} approach is a \emph{continuous time} approach
implemented by adoption of a \emph{stochastic differential equation}, i.e.,
an equation for quantities which are \emph{random} in nature. In other
words, Langevin provided an elegant solution to the problem of generalizing
a dynamical equation to a probabilistic equation. This was to be the start
of a major new field of study with applications in physics, chemistry,
biology, and many other fields.

The approach of Langevin was phenomenological but its essential correctness
has been verified by various microscopic studies. A key feature of his
approach was to separate the total force acting on a particle due to its
environment into two parts: a mean frictional force and a fluctuation
(random) force. These forces are related to each other as a consequence of
the requirement that the equilibrium state be stationary. On the other hand
these two forces are very different in nature: The fluctuation force is
basically microscopic in nature and has a time scale determined by the mean
time between collisions, whereas the time scale of the frictional force is
proportional to the self-diffusion constant and is much larger. Explicitly,
Langevin explained the motion of a "
Brownian," particle (an otherwise free particle in a
dissipative environment) by an elegant stochastic classical differential
equation \cite{langevin08}:
\begin{equation}
m\ddot{x}+\zeta \dot{x}=F(t),  \label{rr1}
\end{equation}
where $m$ and $x$ denote the mass and coordinate of the particle,
respectively, and the dot denotes differentiation with respect to time. The
force on the particle consists of the frictional (dissipative) term $-\zeta 
\dot{x}$ and the random (fluctuation or noise) term $F(t)$. The latter term
is zero at zero temperature, in contrast to the corresponding situation in
quantum mechanics, as we shall see later.

Since the past motion does not appear in the Langevin equation (\ref{rr1}),
one says that there is no memory or, equivalently (in the classical case
under discussion but not in the quantum case to be discussed later), we say
that the process is Markovian. In addition, the autocorrelation of the
random force is a $\delta $ function (a manifestation of the Markovian
process) and is also proportional to $\zeta $. The latter result is a
manifestation of the fluctuation-dissipation theorem.

In addition, with the usual definition of the diffusion constant,
\begin{equation}
D\equiv \frac{1}{2}\lim_{t\rightarrow \infty }\dot{s}(t),  \label{rr2}
\end{equation}
where
\begin{equation}
s(t-t^{\prime })=\left\langle [x(t)-x(t^{\prime })]^{2}\right\rangle
\label{rr3}
\end{equation}
is the mean-square displacement and the angular brackets denote the average
with respect to the canonical ensemble of the system, Langevin readily
obtained the famous Einstein relation
\begin{equation}
D=\frac{kT}{\zeta }.  \label{rr4}
\end{equation}

As noted above, this is an example of an intimate connection between
fluctuations and dissipation. Another example was provided by Nyquist who
showed that the random fluctuations in voltage across a resistor measured by
Johnson are determined by its impedance (the famous Johnson-Nyquist noise 
\cite{nyquist28} in electrical circuits). A general quantum formulation of
the fluctuation-dissipation theorem first appeared in the well known article
of Callen and Welton \cite{callen88}.

We turn now to another key property of the Langevin equation: Whereas the
original Brownian motion experiment and the analysis of it pertained to a
system in thermal equilibrium, the Langevin equation can be generalized in a
simple manner to include an external force on the right hand side and thus
can describe the irreversible approach to equilibrium.

The Langevin equation is phenomenological. As noted above, its derivation
from a quantum microscopic theory was first given by Ford, Kac and Mazur \cite{ford65}.
Then in 1986, in an article entitled "Quantum Langevin
Equation\textquotedblright , \cite{ford88} together with J. T. Lewis we gave a detailed discussion and
presented the general form of the equation consistent with fundamental
physical requirements, in particular causality and the second law of
thermodynamics. Based on these conclusions, we were able to obtain the most general quantum Langevin equation [given below in (\ref{rr6})] for the macroscopic description of a quantum particle with passive dissipation and moving in an arbitrary external potential.  We then showed that the most general form can be realized by a simple oscillator model of a heat bath.  For the purpose of this article, it is convenient to use this model, the so-called independent-oscillator
(IO) model, for which the quantum Hamiltonian is
\begin{equation}
H_{\text{IO}}=\frac{p^{2}}{2m}+V(x)+\sum_{j}\left( \frac{p_{j}^{2}}{2m_{j}}+
\frac{1}{2}m_{j}\omega _{j}^{2}\left( q_{j}-x\right) ^{2}\right) -xf\left(
t\right) .  \label{rr5}
\end{equation}
Here, $V(x)$ is a particle potential energy, $f\left( t\right) $ is an
external applied force, while $x$ and $p$ are the particle position and
momentum operators and $q_{j}$ and $p_{j}$ are those for the $j$'th bath
oscillator. The parameters are $m$, the particle mass, and $m_{j}$ and $
\omega _{j}$, the mass and frequency of the $j$'th bath oscillator.

The procedure used in obtaining the corresponding Langevin equation from the
microscopic Hamiltonian is common to all such problems and consists of use
of the Heisenberg equations of motion to obtain the equations of motion for
both the dynamical variables of the particle ($x,p$) and the dynamical
variables of the heat bath $(q_{j},p_{j})$ \cite{ford88}. These are coupled
equations and the next step is to eliminate the bath variables. This leads
to an inhomogeneous differential equation for the $q_{j}$. Then, typical of
the way that the time-reversal invariance of the original equations is
broken in macroscopic equations, one chooses the retarded solution of this
equation. This solution for $q_{j}$ is then substituted into the equation
for $x$ to get the Langevin equation for a quantum particle of mass $m$
moving in a one dimensional potential $V(x)$ in an arbitrary heat bath and
temperature $T$:
\begin{equation}
m\ddot{x}+\int_{-\infty }^{t}dt_{1}\mu (t-t_{1})\dot{x}(t_{1})+V^{\prime
}(x)=F(t)+f(t),  \label{rr6}
\end{equation}
where the dot and prime denote, respectively, the derivative with respect to 
$t$ and $x$. This is a Heisenberg equation of motion for the coordinate
operator $x$. The coupling with the heat bath is described by two terms: an
operator-valued random force $F(t)$ with mean zero, and a mean force
characterized by the memory function $\mu (t)$. These quantities are given
in terms of the heat bath variables:
\begin{equation}
\mu (t)=\sum_{j}m_{j}\omega _{j}^{2}\cos (\omega _{j}t)\Theta (t),
\label{rr7}
\end{equation}
where $\Theta (t)$ is the Heaviside step function (by convention the memory
function vanishes for negative times), and
\begin{equation}
F(t)=\sum_{j}m_{j}\omega _{j}^{2}q_{j}^{h}(t),  \label{rr8}
\end{equation}
where $q_{j}^{h}(t)$ denotes the general solution of the homogeneous
equation for the heat bath oscillators (corresponding to no interaction).
Using these results, we find that the (symmetric) autocorrelation of $F(t)$
is
\begin{equation}
{\frac{1}{2}}\left\langle F(t)F(t^{\prime })+F(t^{\prime })F(t)\right\rangle
={\frac{1}{\pi }}\int_{0}^{\infty }d\omega \mathrm{Re}\left[ \tilde{\mu}
(\omega +i0^{+})\right] \hbar \omega \coth {\frac{\hbar \omega }{2kT}}\cos
[\omega (t-t^{\prime })],  \label{rr9}
\end{equation}
and the nonequal-time commutator of $F(t)$ is
\begin{equation}
\lbrack F(t),F(t^{\prime })]={\frac{2}{i\pi }}\int_{0}^{\infty }d\omega 
\mathrm{Re}\left[ \tilde{\mu}(\omega +i0^{+})\right] \hbar \omega \sin
[\omega (t-t^{\prime })].  \label{rr10}
\end{equation}
In these expressions
\begin{equation}
\tilde{\mu}(z)=\int_{0}^{\infty }dte^{izt}\mu (t),\quad \mathrm{Im}z>0 ,
\label{rr11}
\end{equation}
is the Fourier transform of the memory function $\mu (t)$. Finally, $F(t)$
has the Gaussian property: correlations of an odd number of factors of $F$
vanish; those of an even number of factors are equal to the sum of products
of pair correlations, the sum being over all pairings with the order of the
factors preserved within each pair.  Equation (\ref{rr9}) is an exact fluctuation-dissipation theorem and we emphasize that it is independent of both the potential $V(x)$ and the external force $f(t)$.  Physically, it expresses the fact that the spontaneous \underline{equilibrium} fluctuations of the heat bath (described by the left side of (\ref{rr9})) are related to the dissipation parameter $\textnormal{Re}\tilde{\mu}(\omega )$.

From the explicit expression (\ref{rr7}) it is clear that the memory
function is dependent only on the bath parameters. We obtained a key
constraint on this function by considering the effect of an arbitrary $c$
-number external force $f(t)$ acting on an otherwise free particle ($V\left(
x\right) =0$) \cite{ford88}. If we assume that $f\left( t\right) $ vanishes
in the distant past and future, the effect is to carry the system of free
particle coupled to the bath in a complete cycle from a state of
equilibrium, through a continuous sequence of intermediate states, and back
to equilibrium. The second law of thermodynamics, in the Kelvin-Planck form,
then requires that the net work done by this force be positive, which in
turn requires that the spectral distribution $\mathrm{Re}\left[ \tilde{\mu}
\left( \omega +i0^{+}\right) \right] $ must satisfy the positivity condition:
\begin{equation}
\mathrm{Re}\left\{ \tilde{\mu}\left( \omega +i0^{+}\right) \right\} \geq
0,\quad -\infty <\omega <\infty .  \label{rr12}
\end{equation}
We note that since the memory function is real the spectral distribution
must also satisfy the reality condition: $\mathrm{Re}\left[ \tilde{\mu}
\left( -\omega +i0^{+}\right) \right] =\mathrm{Re}\left[ \tilde{\mu}\left(
\omega +i0^{+}\right) \right] $.

The positivity condition (\ref{rr12}) together with the fact, obvious from
the definition (\ref{rr11}), that $\tilde{\mu}(z)$ is analytic in the upper
half plane, means that means that $\tilde{\mu}(z)$ must be what is called a
positive real function. This is a very restricted class of functions of a
complex variable, with special properties which include the Stieltjes
inversion theorem:
\begin{equation}
\tilde{\mu}(z)=\frac{2iz}{\pi }\int_{0}^{\infty }d\omega \frac{\mathrm{Re}
\left\{ \tilde{\mu}\left( \omega +i0^{+}\right) \right\} }{z^{2}-\omega ^{2}}
.  \label{rr13}
\end{equation}
Thus, we see that in the quantum Langevin equation the memory function as
well as the correlation and commutator of the random force, are completely
characterized by the spectral distribution.

There are further conditions on the spectral function. As a consequence of
the inversion theorem (\ref{rr13}) we see that
\begin{equation}
\int_{0}^{\infty }d\omega \frac{\mathrm{Re}\left\{ \tilde{\mu}\left( \omega
+i0^{+}\right) \right\} }{1+\omega ^{2}}<\infty .  \label{rr14}
\end{equation}
An important further constraint is what has been called the zero'th law of
thermodynamics: there must be an equilibrium state. This requires that the
spectral distribution must be everywhere positive, with no gaps in which it
vanishes. This in turn requires for our model that the bath frequencies must
be infinite in number and continuously distributed.

In the case of an oscillator potential $V(x)=\frac{1}{2}Kx^{2}$, the
solution of the quantum Langevin equation (\ref{rr6}) is given by \cite
{ford85}
\begin{equation}
\tilde{x}(\omega )=\alpha (\omega )\left[\tilde{F}(\omega )+\tilde{f}(\omega )\right],  \label{rr15}
\end{equation}
where the superposed tilde to denotes the Fourier transform, e.g., $\tilde{x}
(\omega )$ is the Fourier transform of the operator $x(t)$, and $\alpha
(\omega )$ is the generalized susceptibility (a $c$-number) given by
\begin{equation}
\alpha (\omega )=\frac{1}{-m\omega ^{2}-i\omega \tilde{\mu}(\omega )+K}.
\label{rr16}
\end{equation}
As a simple application of this result, we find for the autocorrelation and
commutator of the operator $x(t)$, in linear response, are \cite{ford85,li93}
\begin{eqnarray}
\frac{1}{2}\left\langle x(t)x(t^{\prime })+x(t^{\prime })x(t)\right\rangle
&=&\frac{\hbar }{\pi }\int_{0}^{\infty }d\omega \mathrm{Im}\left\{ \alpha
\left( \omega +i0^{+}\right) \right\} \coth \frac{\hbar \omega }{2kT}\cos
\left( \omega \left( t-t^{\prime }\right) \right) ,  \notag \\
\lbrack x(t),x(t^{\prime })] &=&\frac{2\hbar }{i\pi }\int_{0}^{\infty
}d\omega \mathrm{Im}\{\alpha (\omega +i0^{+})\}\sin \omega (t-t^{\prime }).
\label{rr17}
\end{eqnarray}

Equations (\ref{rr9}), (\ref{rr10}) and (\ref{rr17}) are exact fluctuation-dissipation theorems for our general analysis and they provide the foundation for subsequent developments.  We note, however, that equation (\ref{rr17}) depends on both the potential and the dissipative parameter.  In fact, in the classical case (but not in the quantum case), we showed that, in the case of a weak applied force, the spontaneous fluctuations (described by the left-side of (\ref{rr17})) relax with the same time constant as the induced (by $f(t)$) non-equilibrium fluctuations \cite{ford962}.  This is the essence of the Onsager regression hypothesis \cite{onsager31} which states that regression of fluctuations is governed by macroscopic equations describing the approach to equilibrium.

We have now essentially all the tools that we need and thus it is time to
turn to specific applications. However, here we decided to concentrate on important problems in electrodynamics.  First, in section II, we survey the vast amount of work on the problem of runaway solutions of the Abraham-Lorentz equation for a radiating electron.  Many of these attempts started with force equations which do not encapsulate the important time development nature of the problem.  By contrast, our starting point is a Hamiltonian which beautifully leads to an equation of motion valid for all times.  Moreover, they treated the acceleration $\ddot{x}(t)$ as the key parameter, ignoring the fact that it is not an observable, in contrast to the applied external force $f(t)$ (which, of course, differs from $M\ddot{x}(t)$ because of the dissipation), which is the basic parameter used in our approach.  Since all previous solutions suffered from one problem or another, in section III we present our consistent theory, which led us to a simple second-order differential equation which is not only free of runaway solutions but has no causality problems and is consistent with the optical theorem. Since the basic framework is the same, we also include in this section our treatment of temperature effects on atomic energy and free energy levels.

In section IV, we present the relativistic extension of our theory and in section V we discuss associated fluctuation and quantum effects.  Then, in section VI, we briefly enumerate other applications of our general theory which embraces both radiation reaction and dissipation.  Our conclusions are summarized in section VII.

\section{Radiation reaction in electrodynamics: Historical survey}

The earliest work on radiation reaction is that of Abraham \cite{abraham02}
and Lorentz \cite{lorentz16}, whose result is summarized in the well-known
equation: 
\begin{equation}
M\ddot{x}-M\tau _{e}\dddot{x}=f(t)\quad \text{Abraham-Lorentz},
\label{rr18}
\end{equation}
where
\begin{equation}
\tau _{e}=\frac{2e^{2}}{3Mc^{3}}\simeq 6.25\times 10^{-24}s.  \label{rr19}
\end{equation}
For a careful discussion of the derivation of this equation, which is
"exact," for a point electron, see Jackson 
\cite{jackson98}. This equation exhibits the well-known problem of runaway
solutions: even a small impulsive force acting on an electron at rest
results in an exponentially growing displacement. This is made more explicit in the sentence following equation (\ref{rr45}).  In fact, when $f(t)=0$, equation (\ref{rr18}) does not reduce to Newton's equation, as it should.  In essence, the problem with the equation might be thought to lie in its derivation which was based on force equations, as distinct from a Hamiltonian.  However, as we shall later point out, our exact Hamiltonian approach leads, in the case of a point electron to the same equation (\ref{rr18}), making clear that the basic assumption of a point electron is at fault.  In fact, both Abraham and Lorentz presented a more systematic discussion by considering both the charge structure of the particle and its self-fields.  However, their derivation contained a variety of assumptions and their final result was an infinite expansion, the leading term corresponding to the point electron result given in (\ref{rr18}).

Attempts to solve this problem have engaged the efforts of a large number of investigators over the
past century. Since, in our view, none of these efforts have been
successful, we will just concentrate on presenting the more prominent.  First, we note that Born and Infeld \cite{born34} attempted to fix the problems associated with the Abraham-Lorentz approach by modifying Maxwell's theory to make it non-linear but, in particular, they encountered problems with quantization.

Dirac \cite{dirac38} attempted to solve the problem by including
advanced solutions in addition to retarded solutions but this effort suffered
from a violation of causality (so that the acceleration at time $t$ depended on the force acting at times earlier that $t$).   An attempt by Ivanenko and Sokolov \cite{iwanenko53} to
replace the Abraham-Lorentz equation by an integro-differential equation of
motion was also flawed by virtue of introducing a violation of causality. Similar remarks apply to the theory
of Wheeler and Feynman \cite{wheeler49}. See also \cite{examplerev,eliezer48,landau87};  we discuss the attempts of Eliezer \cite{eliezer48} and Landau and Lifshitz \cite{landau87} in more detail in section III, in connection with our own results.

\section{A consistent theory of radiation reaction}

The Abraham-Lorentz equation is a result of nineteenth century physics with,
moreover, no notion of fluctuations. Almost without exception, discussions
of radiation reaction in the past century have relied on the same physics.
On the other hand, our contribution has been to recognize that a correct
equation is a quantum Langevin equation for which the heat bath is the
blackbody radiation field, with fluctuations due to the fluctuations of that
field. As we have seen, a consistent derivation of such an equation begins
with a Hamiltonian formulation of the dynamics and the laws of
thermodynamics impose a powerful constraint on that formulation.

For a nonrelativistic electron (charge $=-e$) interacting with the quantum
electrodynamic radiation field the Hamiltonian has the form \cite{ford88}:
\begin{equation}
H_{\text{QED}}=\frac{1}{2m}\left( \mathbf{p}+\frac{e}{c}\mathbf{A}\right)
^{2}+V(\mathbf{r})+\sum_{\mathbf{k},s}\hbar \omega _{k}a_{k,s}^{\dag
}a_{k,s}-xf\left( t\right) ,  \label{rr20}
\end{equation}
where the vector potential is given by
\begin{equation}
\mathbf{A}=\sum_{\mathbf{k},s}\sqrt{{\frac{2\pi \hbar c^{2}}{\omega _{k}V}}}
f_{k}\mathbf{\hat{e}}_{\mathbf{k},s}\left( a_{\mathbf{k},s}+a_{\mathbf{k}
,s}^{\dag }\right) .  \label{rr21}
\end{equation}
Here the symbols have their usual meanings. The quantity $f_{k}$ is the
electron form factor (Fourier transform of the electron charge
distribution). Without loss of generality, we have taken the form factor as
well as the polarization vector $\hat{e}_{k,s}$, to be real. The form
factor, which is sometimes called a cutoff factor, must have the property
that it is unity up to some large cutoff frequency $\Omega $ beyond which it
falls to zero. Then, we showed \cite{ford88} that, by a
unitary transformation that leaves the position operator unchanged, this
Hamiltonian can be put into the independent oscillator form (\ref{rr5}).
Rather than repeating that discussion, we will show how to get the quantum
Langevin equation directly from this form of the Hamiltonian.

The Heisenberg equations of motion are
\begin{eqnarray}
\mathbf{\dot{r}} &=&\frac{1}{i\hbar }\left[ \mathbf{r},H_{\text{QED}}\right]
=\frac{\mathbf{p}+\frac{e}{c}\mathbf{A}}{m},  \notag \\
\mathbf{\dot{p}} &=&\frac{1}{i\hbar }\left[ \mathbf{p},H_{\text{QED}}\right]
=-\frac{\partial V}{\partial \mathbf{r}}+f\left( t\right) ,  \notag \\
\dot{a}_{\mathbf{k},s} &=&\frac{1}{i\hbar }\left[ a_{\mathbf{k},s},H_{\text{
QED}}\right] =-i\omega _{k}-i\sqrt{{\frac{2\pi e^{2}}{\hbar \omega _{k}V}}}
f_{k}\mathbf{\hat{e}}_{\mathbf{k},s}\cdot \frac{\mathbf{p}+\frac{e}{c}
\mathbf{A}}{m}.  \label{rr22}
\end{eqnarray}
Eliminating the particle momentum operator between the first two equations,
we get the particle equation of motion:
\begin{equation}
m\mathbf{\ddot{r}+}\frac{\partial V}{\partial \mathbf{r}}=\frac{e}{c}\mathbf{
\dot{A}.}  \label{rr23}
\end{equation}
The solution of the last equation can be written
\begin{equation}
a_{\mathbf{k},s}\left( t\right) =a_{\mathbf{k},s}^{\left( h\right) }\left(
t\right) -i\sqrt{{\frac{2\pi e^{2}}{\hbar \omega _{k}V}}}f_{k}\mathbf{\hat{e}
}_{\mathbf{k},s}\cdot \int_{-\infty }^{t}dt^{\prime }e^{-i\omega _{k}\left(
t-t^{\prime }\right) }\mathbf{\dot{r}}\left( t^{\prime }\right) ,
\label{rr24}
\end{equation}
where $a_{\mathbf{k},s}^{\left( h\right) }\left( t\right) $ is the solution
of the homogeneous equation, corresponding to free motion of the bath in the
absence of the electron. Putting this in the expression (\ref{rr21}) for the
vector potential, we find
\begin{equation}
\mathbf{A}\left( t\right) =\mathbf{A}^{\left( h\right) }\left( t\right) -
\frac{4\pi ec}{V}\sum_{\mathbf{k},s}\frac{f_{k}^{2}}{\omega _{k}}
\int_{-\infty }^{t}dt^{\prime }\sin \left[ \omega _{k}\left( t-t^{\prime
}\right) \mathbf{\hat{e}}_{\mathbf{k},s}\mathbf{\hat{e}}_{\mathbf{k},s}\cdot 
\mathbf{\dot{r}}\left( t^{\prime }\right) \right] .  \label{rr25}
\end{equation}
The sum over $s$ is the sum over the two polarization directions
perpendicular to $\mathbf{k}$, so we have $\sum_{s}\mathbf{\hat{e}}_{\mathbf{
k},s}\mathbf{\hat{e}}_{\mathbf{k},s}\cdot \mathbf{\dot{r}=\dot{r}}-\mathbf{
\hat{k}}\cdot \mathbf{\dot{r}\hat{k}}$. Next we form the limit of an
infinite quantization volume, using the prescription $\sum_{\mathbf{k}
}\rightarrow \frac{V}{\left( 2\pi \right) ^{3}}\int d\mathbf{k}$. With this
the particle equation of motion (\ref{rr23}) takes the form of a quantum
Langevin equation:
\begin{equation}
m\mathbf{\ddot{r}}+\int_{-\infty }^{t}dt^{\prime }\mu \left( t-t^{\prime
}\right) \mathbf{\dot{r}}\left( t^{\prime }\right) +\frac{\partial V}{
\partial \mathbf{r}}=\mathbf{F}\left( t\right) +f\left( t\right) ,
\label{rr26}
\end{equation}
where the memory function is
\begin{equation}
\mu \left( t\right) =\frac{4e^{2}}{3\pi c^{3}}\int_{0}^{\infty }d\omega
_{k}\omega _{k}^{2}f_{k}^{2}\cos \left( \omega _{k}t\right) \theta \left(
t\right) ,   \label{rr27}
\end{equation}
and the fluctuating operator force is
\begin{equation}
\mathbf{F}\left( t\right) =\frac{e}{c}\mathbf{\dot{A}}^{\left( h\right)
}\left( t\right) .  \label{rr28}
\end{equation}
Note that these expressions for the memory function and the fluctuating
force satisfy the general feature that they depend only on the bath
parameters, independent of the particle mass and the external potential.

From the expression (\ref{rr27}) for the memory function, we find that the
spectral distribution is
\begin{equation}
\mathrm{Re}\left[ \tilde{\mu}\left( \omega +i0^{+}\right) \right] ={\frac{
4\pi e^{2}}{6\pi c^{3}}}\int d\omega _{k}\omega _{k}^{2}f_{k}^{2}\delta
\left( \omega -\omega _{k}\right) ={\frac{2e^{2}\omega ^{2}}{3c^{3}}}
f_{k}^{2}.  \label{rr29}
\end{equation}
The physically significant results for this model should not depend upon
details of the electron form factor, subject, of course, to the condition
that it be unity up to some large frequency $\Omega $ and falls to zero
thereafter. A convenient form which satisfies this condition is
\begin{equation}
f_{k}^{2}=\frac{\Omega ^{2}}{\omega _{k}^{2}+\Omega ^{2}}.  \label{rr30}
\end{equation}
Using this in the expression (\ref{rr29}) for the spectral distribution, the
Stieltjes inversion formula (\ref{rr13}) gives
\begin{equation}
\tilde{\mu}(z)=\frac{2e^{2}}{3c^{3}}\frac{z\Omega ^{2}}{z+i\Omega }.
\label{rr31}
\end{equation}
Finally, with the form (\ref{rr30}) for the form factor, the expression (\ref{rr27}) for the memory function can be evaluated to give
\begin{equation}
\mu \left( t\right) =\frac{2e^{2}}{3c^{3}}\Omega ^{2}\left( 2\delta \left(
t\right) -\Omega e^{-\Omega t}\right) \theta \left( t\right) .  \label{rr32}
\end{equation}

Put this in the equation of motion (note $\delta \left( t\right) \theta
\left( t\right) $ is "half" $\delta (t)$)
with $\left. V\left( \mathbf{r},t\right) =\frac{1}{2}Kr^{2}-\mathbf{r}\cdot 
\mathbf{f}\left( t\right) \right. $, corresponding to an external oscillator
potential and an applied force $\mathbf{f}\left( t\right) $. Then multiply
both sides by $e^{\Omega t}$ and differentiate with respect to $t$, to get
the equation of motion in the form:
\begin{equation}
\frac{1}{\Omega }m\mathbf{\dddot{r}}+\left( m+\frac{2e^{2}}{3c^{3}}\Omega
\right) \mathbf{\ddot{r}}+\frac{1}{\Omega }K\mathbf{\dot{r}}+K\mathbf{r}=
\mathbf{f}\left( t\right) +\frac{1}{\Omega }\mathbf{\dot{f}}\left( t\right) +
\mathbf{F}\left( t\right) +\frac{1}{\Omega }\mathbf{\dot{F}}\left( t\right) .
\label{rr33}
\end{equation}
For motion for which the typical frequencies are much below the cutoff
frequency $\Omega $, this becomes of the form for a free particle, but with
a renormalized mass:
\begin{equation}
M=m+\frac{2e^{2}\Omega }{3c^{3}}.  \label{rr34}
\end{equation}
The mass $M$ is interpreted as the observed mass of the electron, although
at ultrahigh frequencies $m$, the bare mass, reappears. Now with this
interpretation the bare mass, what we have up till now been calling the
electron mass, is no longer observable, but is given in terms of the
observed quantities through
\begin{equation}
m=M\left( 1-\Omega \tau _{e}\right) ,  \label{rr35}
\end{equation}
where $M=9.11\times 10^{-28}g$ and $\tau
_{e}=2e^{2}/3Mc^{3}=6.25\times 10^{-24}s$. With this expression for
the bare mass, the equation (\ref{rr33}) can be written
\begin{equation}
M\left( \frac{1}{\Omega }-\tau _{e}\right) \mathbf{\dddot{r}}+M\mathbf{\ddot{
r}}+\frac{1}{\Omega }K\mathbf{\dot{r}}+K\mathbf{r}=\mathbf{F}\left( t\right)
+\frac{1}{\Omega }\mathbf{\dot{F}}\left( t\right) +\mathbf{f}\left( t\right)
+\frac{1}{\Omega }\mathbf{\dot{f}}\left( t\right) .  \label{rr36}
\end{equation}
Within the framework of our model, this is an exact Heisenberg operator
equation of motion.

For the purpose of making contact with the Abraham-Lorentz and other
classical equations, in Eq. (\ref{rr36}) we will take mean values, set $K=0$
and specialize to one dimension. Thus, the fluctuating force $F(t)$ is
eliminated and all quantities are now to be interpreted as classical
quantities and we obtain
\begin{equation}
M\left( \Omega ^{-1}-\tau _{e}\right) \dddot{x}(t)+M\ddot{x}(t)=f(t)+\Omega
^{-1}\dot{f}(t).  \label{rr37}
\end{equation}
We immediately see that the Abraham-Lorentz equation (\ref{rr18}) follows
if we take $\Omega \rightarrow \infty $, corresponding to a point-electron
model for the electron. But in this limit the bare mass (\ref{rr35}) is
negative infinity. This is the source of the runaway solutions that plague
that model.

If the bare mass is to be positive the relation (\ref{rr35}) puts a
constraint on the cutoff frequency:
\begin{equation}
\Omega \leq \tau _{e}^{-1}  \label{rr38}
\end{equation}
and hence the point electron model, associated with the Abraham-Lorentz
equation, is ruled out. The largest possible value of the cutoff consistent
with this constraint is $\Omega =\tau _{e}^{-1}$, corresponding to zero bare
mass. Choosing this value of the cutoff, the equation of motion (\ref{rr36})
becomes
\begin{equation}
M\mathbf{\ddot{r}}+\tau _{e}K\mathbf{\dot{r}}+K\mathbf{r}=\mathbf{f}\left(
t\right) +\tau _{e}\mathbf{\dot{f}}\left( t\right) +\mathbf{F}\left(
t\right) +\tau _{e}\mathbf{\dot{F}}\left( t\right) .  \label{rr39}
\end{equation}
This is a rather striking exact result in that it is only a second-order
equation with the only parameter being $\tau _{e}$. Its form is a result of
our choice (\ref{rr30}) for the cutoff function with the choice $\Omega
=\tau _{e}^{-1}$. Other forms of the cutoff function will give rise to terms
on the right hand side of higher order in $\tau _{e}$, but the first order
term is the same for all. Since these higher order terms reflect meaningless
details of the cutoff function, we feel that the simple equation (\ref{rr39}) is the one of choice. It is our key result.

The case usually discussed in the literature corresponds to a free classical
particle ($K=0$) with neglect of fluctuations ($F\left( t\right) =0$) and
motion in one dimension, where our equation (\ref{rr39}) specializes to the
form
\begin{equation}
M\ddot{x}=f\left( t\right) +\tau _{e}\dot{f}\left( t\right) .\quad \text{
Ford-O'Connell.}  \label{rr40} 
\end{equation}
We emphasize that $f(t)$ is a general time-dependent external field.  In the particular case where this field is an electric field, we found that \cite{intravaia11,ford913}

\begin{equation}
f(t)=\left(\frac{1}{1+\omega^{2}\tau^{2}_{e}}\right)^{1/2}~~ e~E(t) , \label{rr41}
\end{equation}
which, except for very large frequencies $\omega$ is essentially $e~E(t)$.  It is of interest to note that Eliezer \cite{eliezer48}, whose approach was to postulate what he considered possible solutions, also wrote down (\ref{rr40}) with $f(t)=e~E(t)$ [see Eliezer's equation (\ref{rr9})].  Landau and Lifshitz \cite{landau87} later obtained by a method of successive approximations to the Abraham-Lorentz equation a result similar to that of Eliezer.  Of course, neither of these investigators realized that all departures from the AL equation require the existence of a charge with structure.

Equation (\ref{rr40}) is to be compared with the Abraham-Lorentz equation (\ref
{rr18}). Indeed, if in that equation we assume that in first approximation
the term $-M\tau _{e}\dddot{x}$ is small and can be neglected, we get $M
\ddot{x}\cong f\left( t\right) $ and $-M\tau _{e}\dddot{x}\cong -\tau _{e}
\dot{f}\left( t\right) $, giving our equation (\ref{rr40}). This is
essentially the argument used by  Eliezer \cite
{eliezer48} and by Landau and Lifshitz \cite{landau87} to get a corresponding result. But, due to the existence of
runaway solutions, the term $-M\tau _{e}\dddot{x}$ is never small for all
times, so their argument is flawed. On the other hand, as we have seen, our
equation (\ref{rr40}) is the result of a consistent theory with consistent
approximations.

Forming the Fourier transform of equation (\ref{rr39}), we can write the
solution as
\begin{equation}
\mathbf{\tilde{r}}\left( \omega \right) =\alpha \left( \omega \right) \left[ 
\mathbf{\tilde{f}}\left( \omega \right) +\mathbf{\tilde{F}}\left( \omega
\right) \right] ,  \label{rr42}
\end{equation}
where the polarizability is given by

\begin{equation}
\alpha \left( \omega \right) =\frac{1-i\omega \tau _{e}}{-M\omega
^{2}+\left( 1-i\omega \tau _{e}\right) K}.  \label{rr43}
\end{equation}
As a simple application we can calculate the autocorrelation (\ref{rr17})
in the classical limit ($\hbar \rightarrow 0$). With the polarizability (\ref
{rr43}) we find
\begin{eqnarray}
\frac{1}{2}\left\langle x(t)x(0)+x(0)x(t)\right\rangle  &=&\frac{2kT\tau _{e}
}{M\pi }\int_{0}^{\infty }d\omega \frac{\omega ^{2}}{\left( \omega
_{0}^{2}-\omega ^{2}\right) ^{2}+\gamma ^{2}\omega ^{2}}\cos \left( \omega
t\right)   \notag \\
&=&\frac{kT}{K}e^{-\gamma \left\vert t\right\vert /2}\left[ \cos \left(
\omega _{1}t\right) -\frac{\gamma }{2\omega _{1}}\sin \left( \omega
_{1}\left\vert t\right\vert \right) \right] ,  \label{rr44}
\end{eqnarray}
where
\begin{equation}
\omega _{0}^{2}=\frac{K}{M},\quad \gamma =\frac{K\tau _{e}}{M},\quad \omega
_{1}=\sqrt{\omega _{0}^{2}-\frac{\gamma ^{2}}{4}}.  \label{rr45}
\end{equation}
If we make the same calculation with the Abraham-Lorentz equation we find an
autocorrelation that grows exponentially in time, a clearly unphysical
result which shows that in the presence of fluctuations the problem of
runaway solutions is inescapable.

It is well known that the Abraham-Lorentz equation is compatible with the
Larmor formula for the radiated power. However, for our
equation (\ref{rr40}) we find that the total electromagnetic energy
radiated from a confined current distribution \cite{ford913,ford912} is

\begin{equation}
W_{\text{R}} =\frac{2e^{2}}{3c^{3}}\int_{-\infty }^{\infty }dt\left( \frac{\mathbf{f}
\left( t\right) }{M}\right) ^{2}.  \label{rr46}
\end{equation}
This is our
generalization of the Larmor formula. If we make the replacement $\mathbf{f}
\left( t\right) \rightarrow M\mathbf{\ddot{r}}$, which corresponds to
setting $\tau _{e}=\Omega ^{-1}\rightarrow 0$, we get the familiar form of
the Larmor formula. Thus we see that our expression (\ref{rr46}) for the
radiated energy is compatible with our form (\ref{rr40}) of the equation of
motion, just as the Larmor expression is compatible with the Abraham-Lorentz
equation. Finally, we should emphasize that for a force $\mathbf{f}\left(
t\right) $ that is slowly varying on a time scale $\tau _{e}$ the difference
between the two expressions is negligibly small but our expression is in terms of the aplied force, as distinct from the theoretically derived acceleration.

At first, one might be surprised that (\ref{rr39}) predicts that for a free
particle ($K=0$) there is no radiation if the external force is constant
whereas (\ref{rr46}) seems to state otherwise. The answer is that (\ref{rr46}) was derived, as is usual, with the assumption that $f(t)$ is
switched on in the distant past and off in the distant future. For a force
that is switched on, is constant for a long time, and then switched off,
there is no radiation during the intermediate times of constant force \cite
{ford912}, yet the total radiated energy is correctly given by the formula (\ref{rr46}). This was verified in an explicit example \cite{ford93}. With
some exceptions, it appears that this simple explanation was often missed in
the endless debate surrounding this problem in the past (where, in dealing
with the Abraham-Lorentz equation, the argument centred on constant
acceleration). The most notable exception was Feynman \cite{feynman03} who
states that "- - we have inherited a prejudice that an
accelerating charge should radiate -- the power radiated by an accelerating
charge [the Larmor formula] has led us astray," and he then
goes on to discuss the limited validity of the Larmor formula and the fact
that "- - it does not suffice to tell us "
when," the energy is radiated\textquotedblright .

Finally, we point out that our equation (\ref{rr39}) is not only free of
runaway solutions but is also consistent with the optical theorem and the
standard formulas for the Rayleigh and Thomson scattering cross sections 
\cite{intravaia11}. Moreover, the corresponding polarizability (\ref{rr43}),
since it is analytic in the upper half $\omega $ plane, is consistent with
the basic physical requirement of causality. 

It is also interesting to note that in our first paper in this general area \cite{ford85}, we considered the motion of a
charged particle in a radiation field, with the purpose of finding the
effect of temperature on atomic energy levels (and, of course, this was the
framework we later used for the treatment of radiation from the electron).
This followed much earlier work by Knight \cite{knight72} at a time when it was considered that the effects were so small as to be unobservable.  However, using high-precision laser spectroscopic techniques, in a remarkable paper, Hollberg and Hall \cite{hollberg84} were able to measure fractional shifts of $\sim 2\times 10^{-12}$.  In our approach, since temperature was involved, we were obliged to consider thermodynamics
and, in particular, free energy. As a result, we obtained a striking exact
result for the free energy of a quantum oscillator interacting with a
blackbody radiation field, which we used to obtain agreement \cite{ford87}
with the experimental results obtained for the energy shifts due to
temperature \cite{hollberg84}. Later, when a flurry of paper appeared
claiming the laws of thermodynamics were invalid in the quantum arena, we
were able to show, using this same free energy result, that these claims
were in fact incorrect \cite{ford05}.

\section{Radiation Reaction: Relativistic Theory}

Analogous to Dirac's extension of the Abraham-Lorentz equation \cite{dirac38},
we have proposed an extension of our equation (\ref{rr39}) to the
relativistic domain \cite{ford93}. The form is
\begin{equation}
M\frac{du^{\mu }}{d\tau }=\frac{e}{c}F_{\;\kappa }^{\mu }u^{\kappa }+\tau
_{e}\frac{e}{c}\left( \frac{d}{d\tau }F_{\;\lambda }^{\mu }u^{\lambda }-
\frac{1}{c^{2}}u^{\mu }u^{\kappa }\frac{d}{d\tau }F_{\kappa \lambda
}u^{\lambda }\right) ,  \label{rr47}
\end{equation}
where $F^{\mu \nu }$ is the external electromagnetic field tensor and

\begin{equation}
u^{\mu }=\frac{dx^{\mu }}{d\tau },\quad d\tau =\frac{1}{c}\sqrt{g_{\mu \nu
}dx^{\mu }dx^{\nu }}=\sqrt{1-\frac{v^{2}}{c^{2}}}dt.  \label{rr48}
\end{equation}
This equation is consistent with the constraint

\begin{equation}
g_{\mu \nu }u^{\mu }u^{\nu }=c^{2},  \label{rr49}
\end{equation}
and in the nonrelativistic limit ($c\rightarrow \infty )$ reduces to our
equation (\ref{rr39}) for the special case of a free particle ($K=0$) and
for the mean motion with no fluctuation ($F\left( t\right) =0$). Also in
that limit, the applied force is $\mathbf{f}\left( t\right) =e\mathbf{E}(t)$
, which tells us that the equation is really only valid for spatially
uniform (although possibly time dependent) fields.

It is also of interest to note that our equation of motion (\ref{rr47}) can
be written in the three-vector form
\begin{equation}
M\frac{d\gamma \mathbf{v}}{dt}=\mathbf{F}+\tau _{e}\left[ \gamma \frac{d
\mathbf{F}}{dt}-\frac{\gamma ^{3}}{c^{2}}\left( \frac{d\mathbf{v}}{dt}\times
(\mathbf{v}\times \mathbf{F})\right) \right] .  \label{rr50}
\end{equation}
Here $\mathbf{v=\dot{r}}$, $\gamma =\sqrt{1-v^{2}/c^{2}}$ and

\begin{equation}
\mathbf{F}=e\left( \mathbf{E}+\mathbf{v}\times \mathbf{B}\right) ,
\label{rr51}
\end{equation}
is the Lorentz force with $\mathbf{E}$ and $\mathbf{B}$ the electric and
magnetic fields. We note that the corrections to the non-relativistic
results are of order $(v/c)^{2}$, as one might expect.

In \cite{ford93} we presented an exact solution of this equation
for the case of an electron travelling between the plates of a
parallel-plate capacitor, for which the electric field is uniform between
the plates and zero otherwise. The result shows explicitly the radiation
occurs only as the electron enters and leaves the field.

\section{Fluctuation and quantum effects}

Earlier discussions of radiation reaction were generally based on classical
electrodynamics and implicitly assumed that fluctuations could be neglected.
For the most part, our discussion above was equivalent with such a classical
description, with the fluctuation force eliminated by taking mean values. As
we have seen in our example (\ref{rr44}), in the classical limit this
corresponds to zero temperature, which is not a serious limitation for the
classical theory. However, in the quantum theory zero point fluctuations are
always present. As an example of the importance of quantum fluctuations, we
consider the commutator of the position and velocity. Forming the commutator
of $\mathbf{r}$ with the first of equations (\ref{rr22}), using the
canonical commutation we find when specialized to one dimension, $[x(t),\dot{
x}(t)]=i\hbar /m$, where $m$ is the bare mass. However, using the expression
(\ref{rr43}) for the polarizability, the formula (\ref{rr17}) for the
non-equal time commutator can be readily evaluated and it is obvious that
only the renormalized mass appears. In particular, in the limit as $
t^{\prime }$ approaches $t$ from above or below we find that $[x(t),\dot{x}
(t\pm 0^{+})]=i\frac{\hbar }{M}\left( 1-\gamma \tau _{e}\right) $, with $M$
the renormalized mass. For a detailed discussion see \cite{ford89}.

As another example of a quantum fluctuation phenomenon, consider the mean
square displacement of a harmonically bound electron at zero temperature,
obtained by setting $t^{\prime }=t$ in the correlation (\ref{rr17}) and
forming the zero temperature limit. With the expression (\ref{rr43}) for $
\alpha \left( \omega \right) $, we obtain an expression in terms of an
integral that is logarithmically divergent. However, inclusion of
retardation would make the integral finite with an upper limit approximately
equal to $Mc^{2}/\hbar $.\cite{ford98b} With this mean square displacement
is found to be

\begin{equation}
\left\langle x^{2}\right\rangle _{T=0}\cong \frac{\hbar }{2M\omega _{0}}
\left( 1+\frac{2\omega _{0}\tau _{e}}{\pi }\log \frac{Mc^{2}}{\hbar \omega
_{0}}\right) .  \label{rr52}
\end{equation}
Since $\omega _{0}\tau _{e}$ is presumed very small this corresponds to a
small increase over the leading factor, which is the mean square width of
the oscillator ground state.

Finally, we remark on quantum tunneling in a dissipative system. For most
systems of interest it was found that dissipation decreases tunneling rates.
However, in the case of the blackbody radiation field, we found that
tunneling actually increased \cite{ford914}. The reason for this exception
to the general rule is the presence of mass renormalization.

\section{Miscellaneous Applications}

The case of constant friction is of special interest. There the spectral
distribution is independent of $\omega $: $\mathrm{Re}\left[ \tilde{\mu}
\left( \omega +i0^{+}\right) \right] =\zeta $, the friction constant. This
is frequently referred to as an Ohmic heat bath. The equation (\ref{rr6})
then takes the form:
\begin{equation}
m\ddot{x}+\zeta \dot{x}+V^{\prime }(x)=F(t).  \label{rr53}
\end{equation}
This is the same form as the original, classical form of the Langevin
equation but here, of course, $x$ and $F$ are operators. In this case, since
the past motion does not appear, one says there is no memory. On the other
hand, the autocorrelation of the quantum mechanical random force \cite
{ford88,ford96} becomes
\begin{eqnarray}
{\frac{1}{2}}\left\langle F(t)F(t^{\prime })+F(t^{\prime })F(t)\right\rangle
&=&{\frac{\zeta }{\pi }}\int_{0}^{\infty }d\omega \hbar \omega \coth {\frac{
\hbar \omega }{2kT}}\cos \left[ \omega \left( t-t^{\prime }\right) \right] 
\notag \\
&=&kT\zeta {\frac{d}{dt}}\coth \left[ \Omega _{\text{T}}(t-t^{\prime })
\right]  \notag \\
&=&kT\zeta \left\{ -\frac{\Omega _{\text{T}}}{\sinh ^{2}\left[ \Omega _{
\text{T}}(t-t^{\prime })\right] }+2\delta \left( t-t^{\prime }\right)
\right\} .  \label{rr54}
\end{eqnarray}
where $\Omega _{\text{T}}=\pi kT/\hbar $. In the limit $\hbar \rightarrow 0$
this becomes the familiar form of classical Brownian motion:
\begin{equation}
\left\langle F(t)F(t^{\prime })\right\rangle \underset{\hbar \rightarrow 0}{
\rightarrow }2kT\zeta \delta (t-t^{\prime }),\quad \text{classical}.
\label{rr55}
\end{equation}
The notion of a Markovian system combines two aspects: a stochastic equation
with no memory, as in (\ref{rr53}), and a delta-function correlation, as in
(\ref{rr55}). The quantum system is never Markovian, in particular the
correlation at zero temperature has no delta-function

The original classical Brownian motion problem corresponds to a free
particle $(K=0)$, spectral distribution independent of frequency $\omega$ and $kT>>\hbar \omega $ (absence of quantum effects), in
which case the position auto-correlation function is readily calculated,
leading to an exact expression for the mean-square displacement, which for
large $t$ reduces to the Einstein result (\ref{rr4}) for the diffusion
constant. However, at low temperatures, non-Markovian quantum effects become
important and cannot be neglected. Anomalous diffusion in quantum Brownian
motion has also attracted much interest as an explanation of various
experiments \cite{ford06}.

Another application of interest relates to Josephson junctions. At first
glance one might be puzzled as to how the quantum Langevin equation applies
in this case. Actually, although we have used the language of particle
motion in our formulation of this description, it should be clear that the
description is more general than the language. Thus, the operator $x$ in the
quantum Langevin equation (\ref{rr6}) can be a generalized displacement
operator. By this we mean an operator $x$ such that a term $V(x,t)=-xf(t)$,
with $c$-number $f(t)$, added to the microscopic Hamiltonian of the system,
results in an added term $f(t)$ on the right-hand side of the equation of
motion. One can therefore apply this description to an equation which is
formally similar to the Langevin equation but in which the physical meaning
of $x$ is different. One must, however, be cautious to check the above
generalized displacement property. It turns out that, for Josephson
junctions, the phase difference $\phi $ of the superconducting wave function
across the junction is such a generalized displacement coordinate. This
enabled us to obtain an expression for the power spectrum of the phase
fluctuations (generalizing an earlier weak coupling limit result of
Josephson) as well the mean square deviation of the phase and the power
spectrum of the voltage fluctuations \cite{ford883}.

Turning to the study of small tunnel junctions, here too the quantum
Langevin equation proved to be an effective tool in determining the
mean-square charge fluctuation on the junction by treating the charge
fluctuation as a generalized coordinate \cite{hu92}. The quantum Langevin
equation has also been used to study quantum transport for a many-body
system \cite{hu87}, the advantage being that the separation of frictional
(dissipative) and random (fluctuating) forces gives a natural separation
between the conductivity and the noise.

In general, the separation of fluctuations (noise) and dissipation in the quantum Langevin equation (\ref{rr6}), together with the fluctuation-dissipation theorems, enables us to systematically analyze noise in many different systems.  In particular, using (\ref{rr17}), we were able to calculate the power spectrum of the coordinate fluctuations in a universal model which we presented for the detection of noise in gravitational wave detector systems \cite{ford012}.

In addition, we found that quantum effects required modification of the
famous Onsager (classical) regression hypothesis \cite{ford962}, which
states that the regression of fluctuations is governed by the macroscopic
equations describing the approach to equilibrium. In other words, the
so-called quantum regression theorem is only correct if one makes various
approximations \cite{ford00}. In fact, we claim that the correct
generalization of the Onsager regression hypothesis is the
fluctuation-dissipation theorem of Callen and Welton \cite{ford962}.

It is also of interest to note that the quantum Langevin equation has been
used in a very straightforward manner to analyze the so-called Unruh
radiation problem. Its use in this context was first considered by Sciama
and co-workers \cite{raine91-31} but they used various approximations which
left their work open to criticism. However, we carried out an exact analysis
of an oscillator (the detector) moving under a constant force with respect
to zero-temperature vacuum and coupled to a one-dimensional scalar field 
\cite{ford063}. We showed that, contrary to the conclusions reached by
Unruh, this system does not radiate despite the fact that it thermalizes at
the Unruh temperature.

More recently, the quantum Langevin equation has been used in the general
area of mesoscopic systems and quantum information. Its use enabled us to
incorporate "entanglement at all times\textquotedblright ,
in contrast to the often used but more approximate master equations. In
particular, it enabled us to analyze exactly the decoherence of a Schr\"{o}dinger cat superposition of Gaussian states, one striking result being the
conclusion that decoherence can occur even in the absence of dissipation 
\cite{ford01}, a result which does not emerge from master equation
calculations. It should be noted that one finds many different master equations in the literature.  However, even for what is referred to as the "exact master equation" (which are exact only in the sense that they incorporate time-dependent coefficients) \cite{hu922}, we presented what we feel is the most transparant exact solution \cite{ford013,ford07}.  Our approach started with the Wigner function equivalent equation since the Wigner function provides the same information as the corresponding density matrix while making the calculations simpler and more transparent.  One striking result we obtained is that the exact master equation problem is equivalent to the Langevin equation for the initial value problem, which was much easier to solve.  We found that serious divergences arose at low temperatures \cite{ford013} and that, even in the high temperature regime \cite{ford-oconnell}, problems also exist, notably the fact that the density matrix is not necessarily positive.  In addition, Karrlein and Grabert \cite{hu922} showed that there is no unique master equation, which is connected with the fact that the Onsager regression hypothesis fails in the quantum regime.  By contrast, our "entanglement at all times" approach, based on the use of the quantum Langevin equation, has none of these problems \cite{ford014}.

Since entanglement is the essence of various schemes to build
small devices for quantum information applications (cryptography, quantum
computing, etc.), we used the quantum Langevin equation to examine how
disentanglement occurs in two-body systems.  In particular, we
found that if the temperature is larger than a critical value that
disentanglement can occur in the absence of dissipation \cite{ford10}. However, the effect
of temperature is very different for disentanglement than for decoherence
because the effect is constant for all time whereas for decoherence the
effect increases with time.

\section{Conclusions}

We have seen that radiation occurs when a system interacts with a heat bath. It
is, in essence, a dissipative term (which is not invariant under $
t\rightarrow -t$) which occurs in the equation of motion of the system. More
generally, it occurs in conjunction with fluctuations and the relation
between these quantities is expressed in the famous quantum
fluctuation-dissipation theorem. In the absence of a heat bath, we still
have quantum fluctuations. However, even in the classical case, we have
temperatures fluctuations. It is only in the classical case where
temperature is neglected do we have zero fluctuations so that the
dissipative (radiation reaction) term only appears. This is mainly the arena
for discussing radiation effects in electrodynamics. More generally, as we
saw above, there are many systems for which fluctuations play a key role in
addition to radiation reaction.

We should remark that we have chosen to write the autocorrelation functions
in their symmetric form since we found that this choice led to the most
elegant presentation. However, clearly other choices are possible,
corresponding to the normal and anti-normal ordering of the $a^{+}$
(creation) and $a$ (annihilation) operators, but it is clear that in all
cases the physical implications are the same, a point emphasized by Milonni
in his detailed discussion of QED effects \cite{milonni93}.

In our discussion of radiation effects in electrodynamics, we have used as our starting point the universally accepted Hamiltonian of quantum electrodynamics.  Treating the blackbody radiation field as a heat bath, this enabled us to obtain an exact equation of motion of an electron in the form of a quantum Langevin equation, in which both dissipation and fluctuations appear.  Imposing the requirement that the second law of thermodynamics must be satisfied leads to an equation of motion that is second order in time with no runaway solutions. Thus our equation, in contrast with the well-known Abraham-Lorentz equation, is consistent with physical principles.  In particular, causality is preserved at all times and the optical theorem is satisfied.

\section*{ACKNOWLEDGMENT}

The author is pleased to acknowledge a longtime collaboration with G. W. Ford (one of the pioneers of this area; see, in particular, \cite{ford65}) which led to the essential results presented here. This work was partially supported by the National Science Foundation under
Grant No. ECCS-1125675. \newline

\noindent \textbf{R. F. O'Connell} received a B.Sc. degree (1953) from the
National University of Ireland (after which he worked for 4 years as a
Telecommunications Engineer), a Ph.D. from the University of Notre Dame
(1962), and a D.Sc. (1975) from the National University of Ireland. Before
joining Louisiana State University (LSU) in 1964, he worked at the Dublin
Institute for Advanced Studies and at IBM. He is presently Boyd Professor
(the highest rank in the LSU System) and Professor of Physics at LSU.
O'Connell was a NAS-NRC Senior Research Associate at the NASA Institute for
Space Studies, New York (1996-1998). He was a visiting scientist at Oak
Ridge National Laboratory (1965); Cambridge University (1970); Montana State
U. (1971); Lawrence Livermore Laboratory (1973); U. of Trondheim (1975); U.
of Paris XI (1975); ICTP Trieste (1976); Laboratoire Aim\'{e} Cotton France
(1983); Universidade Federal De Santa Catarina, Florianopolis, Brazil (1987)
and the Technical University of Denmark (1994). He was also an SRC Senior
Visiting Fellow at the Universities of Oxford and London in 1975-1976. He
spent numerous sojourns as a visiting scientist at the Dublin Institute of
Advanced Studies, the Max-Planck Institute for Quantum Optics, Munich and
the University of Ulm, Germany. With Barker, obtained the first correct
general relativistic calculation of spin and orbital precessions in two-body
systems, which was verified by Breton et al. Science 321, 104 (2008). With
Jesse Greenstein and Henry, deduced that the Greenwich white dwarf star has
a magnetic field which is 300 million times larger than the earth's magnetic
field, simultaneously explaining the origin of the famous Minkowski bands.
With Wigner, obtained a new position operator. In recent years, with G. W.
Ford, demonstrated the broad generality of the use of generalized quantum
Langevin techniques for treating dissipative and fluctuation phenomena in
quantum mechanics. He was awarded the Sir J. J. Larmor Prize in physics,
National University of Ireland (1953); an NAS-NRC Fellowship at the
Institute of Space Studies, New York 1966-68; the Distinguished Research
Master award, Louisiana State University, 1975; a Senior Visiting
Fellowship, from Science Research Council (England), at Oxford University
and Queen Mary College, London, January-June, 1976. He was a Board Member,
Physical Review A, January 1995-December 2000 and on the Advisory Panel,
Journal of Physics A since 2006. He has been a Fellow of the American
Physical Society since 1969.

\end{document}